\documentstyle[referee]{mn}
\input{psfig.sty}

\def \be{\begin{eqnarray}}
\def \ee{\end{eqnarray}}

\begin{document}

\title[Discontinuity modes in NS]
{Non-radial oscillation modes as a probe of density discontinuities in neutron stars}
\author[G. Miniutti {\it et al.}]
{G. Miniutti$^1$, J. A. Pons$^{1}$, E. Berti$^2$, L. Gualtieri$^1$, and V.
Ferrari $^1$ \\ $^1$ Dipartimento di Fisica ``G.Marconi",
 Universit\` a di Roma ``La Sapienza"\\
 and Sezione INFN  ROMA1, piazzale Aldo  Moro
 2, I-00185 Roma, Italy \\
$^2$ Department of Physics, Aristotle University of Thessaloniki,
 Thessaloniki 54006, Greece
}

\maketitle

\begin{abstract}
A phase transition occurring in the inner core of a neutron star could be 
associated to a density discontinuity that would affect the frequency 
spectrum of the non-radial oscillation modes in two ways.
Firstly, it would  produce a softening of the
equation of state, leading to more compact equilibrium configurations
and changing the frequency of the fundamental and pressure modes of the neutron
star. Secondly, a new non-zero frequency g-- mode would appear, associated to
each discontinuity. These discontinuity g--modes have  typical frequencies 
larger than those of g--modes previously studied in the literature (thermal, 
core g-- modes, or g--modes due to chemical inhomogeneities in the outer 
layers), and smaller than that of the fundamental mode; therefore  
they should be distinguishable from the other  modes of non radial oscillation.
In this paper we investigate how  high density discontinuities change  
the frequency spectrum of the non-radial oscillations, in the framework of 
the general relativistic 
theory of stellar perturbations.
Our  purpose  is to understand whether  a gravitational signal,  emitted
at the frequencies of the quasi normal modes, may give some clear  
information on the equation of state of the neutron star and, 
in particular, on the
parameters that characterize the density discontinuity. 
We discuss some astrophysical processes that may be associated
to the  excitation of these modes, 
and estimate how much gravitational  energy should the modes convey to produce
a signal detectable by high frequency gravitational detectors.
\end{abstract}

\begin{keywords}
gravitational waves --- stars: oscillations --- relativity --- stars:neutron
\end{keywords}

\section{Introduction}

The equation of state (EOS) of dense matter at supranuclear densities
has been a
long sought-after problem in nuclear physics. Since the first
observational evidences of their existence in the late 60's, neutron stars
(NS) have proved to be a unique astrophysical environment in which very
different fields of  physics, ranging from nuclear physics to particle
physics and general relativity, can be  tested.
Accurate  estimates of some properties of NSs, such as masses and radii,
associated to further information on the thermal history, 
would allow to set constraints on the nature of nuclear forces.
This has been attempted by several means (but so far with limited success),
by correlating multi-wavelength observations with different telescopes,
X-ray, or $\gamma-$ray observatories \cite{TC99,aquilaX1,rxj1856},
or through the study  of  neutrinos from proto-neutron stars
\cite{pns1,pnsq}, or studying QPO properties \cite{OT99,SV99}.

In the next future, a  new observational window will be opened 
by a network of gravitational wave detectors:
the  ground based  interferometers (LIGO, VIRGO, GEO600, TAMA), which
are now in the final stage of
construction or in the commissioning phase,  will explore the frequency
region of 10-1000~Hz, and the space-based interferometer LISA   will enlarge
the observational window to  $10^{-4}-
10^{-1}~$~Hz. In addition,
a  joint panel of European experts have performed an assessment study
for a new generation of interferometric  detectors
with extremely high  sensitivity  in the kHz region
(EURO,EURO-XYLOPHONE), which would provide a very powerful mean
to investigate the physics of neutron stars through gravitational waves.
Since these instruments are currently under study, it is important to envisage
which  sources they would be able to see,
and  what kind of information could be inferred from  the detected signals.
In this paper, we address to one interesting possibility
in this exciting new field: the prospects
of probing phase transitions occurring  in the inner core of neutron stars.
In particular, we shall focus on  phase transitions that produce a density
discontinuity.

Density  discontinuities may  occur
either at high or  low density, depending on the
physical processes  prevailing in the different regions of the star.
At low density,
the characteristic chemical profile in the
outer layers is determined by the history of the star, because shell burning,
flash nuclear burning and accretion phenomena leave layers of
different composition on its surface.  At the interfaces
between different layers, the chemical composition changes abruptly, and if no
significant diffusion is present, the density gradient is well
approximated by a discontinuity.
This  introduces an additional local source of buoyancy which
gives rise to a discontinuity g--mode,
even if no other source of buoyancy
is present, i.e. even in zero temperature NSs.
The properties of the g--modes due to discontinuities in the outer
layers in cold NSs have been studied by 
Finn (1987) and McDermott (1990). Subsequently, Strohmayer (1993)
computed the oscillation
frequencies of two finite-temperature models  with
a density discontinuity associated to the ${}^{56} Fe$ to ${}^{62} Ni$ 
transition. He found that if the discontinuity is
trapped into the crust  no mode can be found
directly associated with it, while if it is located in the fluid ocean
a g--mode appears, and the frequency
distribution of the g--modes due to entropy gradients changes.

However, density discontinuities  are not necessarily
confined to the outer layers of a neutron star: they could also arise as a consequence of
phase transitions in the inner core, where the equation of state is still
poorly known. Several first or second order phase transitions have been
proposed to occur at supranuclear densities,
and they typically involve pion and/or kaon condensation as well
as the transition from ordinary nuclear matter to quark matter
\cite{heise,Pra97}.
Whether these transitions are first or second order, and whether
they admit density discontinuities or a  continuous transition
with a mixed phase, is still matter of debate.
In modeling the phase transition, one 
possibility is  to impose local conservation laws and  to use
a Maxwell construction, which produces a density discontinuity. 
Glendenning (1992) pointed out that, since conservation laws must be imposed
globally,  any first order phase transition  should be accompanied by
a smooth mixed phase region, which satisfies the Gibbs rules.
However, it should be noted that there is much uncertainty on the effect
that Coulomb repulsive forces and surface tension have on the
structure of the mixed phase. If Coulomb and surface energies are large,
the mixed phase is disfavoured,  the region where it occurs shrinks,
and this seems to be the case in a quark-hadron  phase transition \cite{CFL1}.
In this case,  density discontinuities may appear 
near the boundaries of the Gibbs' structured phase,
when the volume fraction of one of the phases is very small.

Astrophysical observations play an important role in clarifying these issues,
because  neutron stars provide a unique opportunity
to study  the behaviour of matter in this very  high density regime.
In the present paper we study  how density discontinuities affect
the spectrum of the quasi normal modes of a neutron star, with the purpose of understanding
what kind of information can be
inferred on the high density equation of state  from
the gravitational signal emitted by an oscillating neutron star.
We shall consider  some astrophysical processes  in which  the
quasi normal modes could be
excited (a glitch,  the onset of the phase transition, binary
coalescence of NS's) and  we shall
estimate the amount of energy that should go into such modes  to produce
a signal that may be detected by the proposed high frequency
gravitational detectors.
In this respect, our work differs from a previous work of
Sotani et al. \cite{sotani}, who focused on
the calculation of the modes and discussed the
stability of the considered models. For simplicity, we consider 
the case of cold neutron stars and  assume a simple polytropic EOS. 

In Sec. 2 we briefly describe the relevant equations of stellar perturbations which we
integrate to  find the  mode frequencies. In Sec. 3 we discuss the results of the
integration, extracting the information that the discontinuity g--modes
carry on the parameters of the density discontinuity. In Sec. 4 we consider
some astrophysical processes that  are associated to the excitation of the quasi-normal
modes, and discuss the detectability of the emitted gravitational radiation.
In Sec. 5 we draw the conclusions.

\section{The mathematical framework}

In order to find the frequencies of the quasi-normal modes,
we will integrate the equations describing the polar, non radial perturbations of a non rotating
star  as formulated by Lindblom and Detweiler \cite{lindet1,lindet2}.
We write the perturbed metric tensor as
\begin{eqnarray}
\label{ds2}
 ds^2 = &-& e^{\nu} (1+r^l H_{0 lm} Y_{lm} e^{i\omega t}) dt^2 - 2 i
        \omega r^{l+1} H_{1 lm} Y_{lm} e^{i\omega t} dt dr
        + e^{\lambda} (1 - r^l H_{0 lm} Y_{lm} e^{i \omega t}) dr^2 \nonumber \\
         &+& r^2 (1 - r^l K_{lm} Y_{lm} e^{i\omega t}) (d\theta^2
        + \sin^2\theta d\phi^2),
\end{eqnarray}
and the polar components of the Lagrangian  displacement of the perturbed
fluid elements as
\begin{eqnarray}
\label{xidef}
 \xi^r &=& e^{-\lambda /2} r^{l-1} W_{lm} Y_{lm} e^{i\omega t} \ ,\nonumber \\
 \xi^\theta &=& -  r^{l-2} V_{lm} \partial_\theta  Y_{lm} e^{i\omega t} \ , \\
 \xi^\phi &=&  - r^l (r\sin\theta)^{-2} V_{lm} \partial_\phi Y_{lm} e^{i\omega t}
  \nonumber \ .
\end{eqnarray}
$Y_{lm}(\theta,\phi)$ are the spherical harmonic functions. In the following we shall
restrict our analysis to the $l= m = 2$ component, which dominates the emission of
gravitational radiation.
By defining the variable $X$
\begin{equation}
\label{Xdef}
 X = \omega^2 (\rho +p)e^{-\nu /2} V - r^{-1} e^{(\nu -\lambda )/2} p^{\,'} W
   +\frac{1}{2} (\rho +p)e^{\nu /2} H_0 \ ,
\end{equation}
where  $p$ and $\rho$ are the fluid's pressure and energy density and
the prime indicates differentiation with respect to $r$,  one can write
a fourth-order system of linear equations for $(H_1 , K , W ,
X)$
\begin{eqnarray}
\label{ldeq}
 H_1^{\,'} &=& \frac{1}{r} \left( -\left[(l+1)+e^{\lambda}
               (2M(r)/r +r^2(p-\rho)) \right] H_1 + e^{\lambda} [ H_0+K
               -4(\rho+p)V ] \right) \ , \nonumber\\
 K^{\,'} &=& \frac{1}{r} \left( H_0 + (n+1)H_1 - [(l+1)-r \frac{1}{2}\,
             \nu^{\,'}] K
          - 2 (\rho+p) e^{\lambda /2} W \right) \ , \nonumber\\
 W^{\,'} &=& -\frac{(l+1)}{r}W + r e^{\lambda /2} \left(\frac{
             e^{-\nu /2} X}{\gamma p} - \frac{l(l+1) V}{r^2} + \frac{1}{2} H_0
             + K \right) \ ,\nonumber \\
 X^{\,'} &=& - \frac{l}{r} X + (\rho+p) e^{\nu /2}
             \left\{\frac{1}{2} \left(\frac{1}{r} - \frac{1}{2} \,
               \nu^{\,'} \right) H_0
             + \frac{1}{2r} [ r^2 \tilde{\omega}^2 + (n+1) ] H_1 +
              \frac{1}{2r} ( \frac{3}{2}\, r \nu^{\,'} -1 )
                K \right. \nonumber \\
              && \left. - \nu^{\,'} \frac{ l(l+1)V}{r^2} -
                \frac{1}{r} \left[ e^{\lambda /2}[(\rho+p)+ \tilde{\omega}^2]
               - \frac{1}{2}\,{r^2} (r^{-2}e^{-\lambda /2}
             \nu^{\,'})^{\,'} \right] W \right\} \ ,
\end{eqnarray}
where $n= (l-1)(l+2)/2,$  $~\tilde{\omega}^2 =\omega^2 e^{-\nu}$,
and $\gamma $ is the adiabatic index
\begin{equation}
\gamma = \frac{\rho +p}{p}\,
      \left( \frac{\partial p}{\partial \rho} \right)_{\rm s}
       = \frac{\rho +p}{p}\,{c^2_{\rm s}} \ .
\end{equation}
The functions $V$ and $H_0$  are
\begin{eqnarray}
\label{elim} H_0 &=& \frac{2 r^2 e^{-\nu /2} X - [\frac{1}{2}\,(n+1)r
\nu^{\,'} - r^2 \tilde{\omega}^2 ] e^{-\lambda} H_1 + [n -
\tilde{\omega}^2 r^2 -\frac{1}{2}\,
\nu^{\,'} (3M(r)-r+r^3p) ] K} {3M(r)/r + n +r^2 p} \nonumber  \\ V &=& \frac{1}{
\tilde{\omega^{2}} } \left( \frac{X e^{-\nu /2}}{(\rho+p)} - \frac{\nu^{\,'}
e^{-\lambda /2} W}{2r} - \frac{1}{2} H_0\right) \ .
\end{eqnarray}
For each given value of $l$ and $\omega$,  Eqs. (\ref{ldeq})
admit only two linearly independent solutions regular at
$r=0$; we find the general solution  as a linear combination of the two, such that
the perturbation of the Lagrangian pressure, $\Delta p$, vanishes at the surface.
This procedure is slightly different from  that proposed by
Lindblom and Detweiler (1983), but  a comparison of the results
of the two procedures shows that they  are absolutely equivalent.

It should be noted that, in this formulation, both the
$\xi^r$  and $\Delta p$  are continuous  even when the density is discontinuous.

It is well known that outside the star, the  variables associated to the  fluid motion
are zero and the equations reduce to a
second-order equation for  the Zerilli function $Z$
\begin{eqnarray}
\label{zer} && \left(\frac{d^2}{dr_*^2} +\omega^2\right) Z =
          U Z \ ,\nonumber \\
&& U= \frac{2(r-2M)}{r^4 (nr+3M)^2}\,
         \left[n^2 (n+1) r^3 +3Mn^2 r^2 +9M^2 (M+nr) \right] \ ,
\end{eqnarray}
where $r_* \equiv r +2M\log (r/2M -1)$ and $M$ is the mass of the star.
The asymptotic behaviour of the Zerilli function is
\begin{equation}
\label{asympt} Z \sim A_{\rm{in}} (\omega ) e^{\rm{i} \omega r_*} +
                      A_{\rm{out}} (\omega ) e^{- \rm{i} \omega r_*}  \ .
\end{equation}
A quasi normal mode of the star is defined to be a solution of the perturbed equations
belonging to a complex eigenfrequency $~\omega_0 + i \omega_{\rm{im}}$, 
which is regular at the center,  continuous at the surface, and which
behaves as a pure outgoing wave at infinity. Therefore,
if $\omega_0$ is the mode frequency,
$A_{\rm{in}} (\omega_0 )$ must vanish.
In order to find the frequency of the quasi-normal modes
we use the following algorithm \cite{CF90}, which is appropriate  when
the real part of the frequency $\omega_0$ is much larger than the imaginary part
$\omega_{\rm{im}}=1/\tau$, as it is in our case.
For real values of the frequency, the function $Z$ must have the asymptotic form
\begin{eqnarray}
\label{Zasympt} Z &\rightarrow&
\left\{\alpha-\frac{n+1}{\omega}\frac{\beta}{r}-
\frac{1}{2\omega^2}\left[n(n+1)\alpha-3M\omega\left(1+\frac{2}{n}\right)
\beta\right]\frac{1}{r^2}+\dots\right\}\cos\omega r_*
\nonumber \\
&-&\left\{\beta+\frac{n+1}{\omega}\frac{\alpha}{r}-
\frac{1}{2\omega^2}\left[n(n+1)\beta+3M\omega\left(1+\frac{2}{n}\right)
\alpha\right]\frac{1}{r^2}+\dots\right\}\sin\omega r_* \ .
\end{eqnarray}
The functions $\alpha (\omega )$ and $\beta (\omega )$ can be  determined by
matching the numerically integrated solution with the above asymptotic
expression for $Z$. The amplitude of the standing gravitational wave at
infinity is $(\alpha^2 +\beta^2 )^{1/2},$ and it can be shown to have a deep
minimum at the frequency $\omega_0$ where  $A_{\rm{in}} (\omega_0 )$ vanishes
\cite{CFW91};
thus, we find the  frequency of a quasi normal mode by searching for  the values 
of $\omega_0$ where $(\alpha^2 +\beta^2 )^{1/2}$ has a minimum.

\section{Results}
To study the effect of high density discontinuities on the oscillation
spectrum of a NS, we shall consider a simple polytropic
equation of state of the form
\begin{equation}
\label{eos}
 p = \left\{
           \begin{array}{rl}
            K\rho^\Gamma \;\;\;\;\;\;\;\;\;\;\;\;\;\;\;\;\;\;\;\;
            &  \;\;\;\;\;\rho > \rho_{\rm{d}} +\Delta\rho \\
            \displaystyle{ K\left( 1 + \frac{\Delta \rho}{\rho_{\rm{d}}}
            \right)^\Gamma \rho^\Gamma}
            & \;\;\;\;\; \rho < \rho_{\rm{d}}
           \end{array}
     \right. \ ,
\end{equation}
where the discontinuity of amplitude $\Delta\rho,$ is located at a density
$\rho_{\rm{d}}$, which we choose to be greater than the nuclear saturation
density $\rho_0 = 2.8\cdot10^{14}$~g/cm$^3$. This simplified equation of
state will allow us to study in a systematic way the properties of
the quasi normal modes, and
the relation between the mode frequencies and the global properties of the
chosen models. This EOS has been used in the past to study the
g--modes associated  to low density discontinuities 
\cite{Finn87,McDermott90} and, in a slightly modified form, to
study the g--modes due to high-density discontinuities (Sotani {\it
et al.}, 2001).

To find the frequency of the surface  g--modes, Finn introduced the
so called slow-motion formalism \cite{Finn86} which is very accurate in
the low-frequency limit. In our case the mode frequencies are
higher since they are  associated to high
density discontinuities, and   the full system of perturbation equations has to be
integrated. As a consistency check of our code, we have repeated some of Finn's
calculations for low density discontinuities \cite{Finn87} finding an excellent
agreement (better than three significant digits).

It should be mentioned that  g-modes may have real, imaginary or zero
frequency, depending on whether the considered stellar model is stable under convection;
different regimes can be identified by looking at  the sign of the   square
of the Brunt-V\"{a}is\"{a}l\"{a} frequency
\begin{equation}
 N^2  = \frac{1}{2} e^{\nu - \lambda } \, \nu^{\,'} \,
                \frac{p^{\,'}}{p} \left(\frac{1}{\gamma} -
                        \frac{1}{\gamma_0} \right)\ , \quad {\mathrm{with}} \quad
                        \gamma_0 = \frac{\rho + p}{p} \, \frac{p^{\,'}}{\rho^{\,'}} \ .
                        \end{equation}
A  real, imaginary or zero  $N$  correspond, respectively, to convective stability,
instability or marginal instability.
If the star is isentropic and chemically homogeneous,  $\gamma = \gamma_0$ and all
g-modes degenerate to zero frequency.
Thus, in the case of our zero temperature, chemically homogeneous star, $N$  is zero everywhere,
except when the density changes abruptly  at  $r=R_{\rm d}$;
at that point, the adiabatic index $\gamma$ remains finite,
the equilibrium index $\gamma_0$ vanishes, $N$  is different from zero,
and a g--mode appears.

\begin{center}
\begin{figure}
\psfig{figure=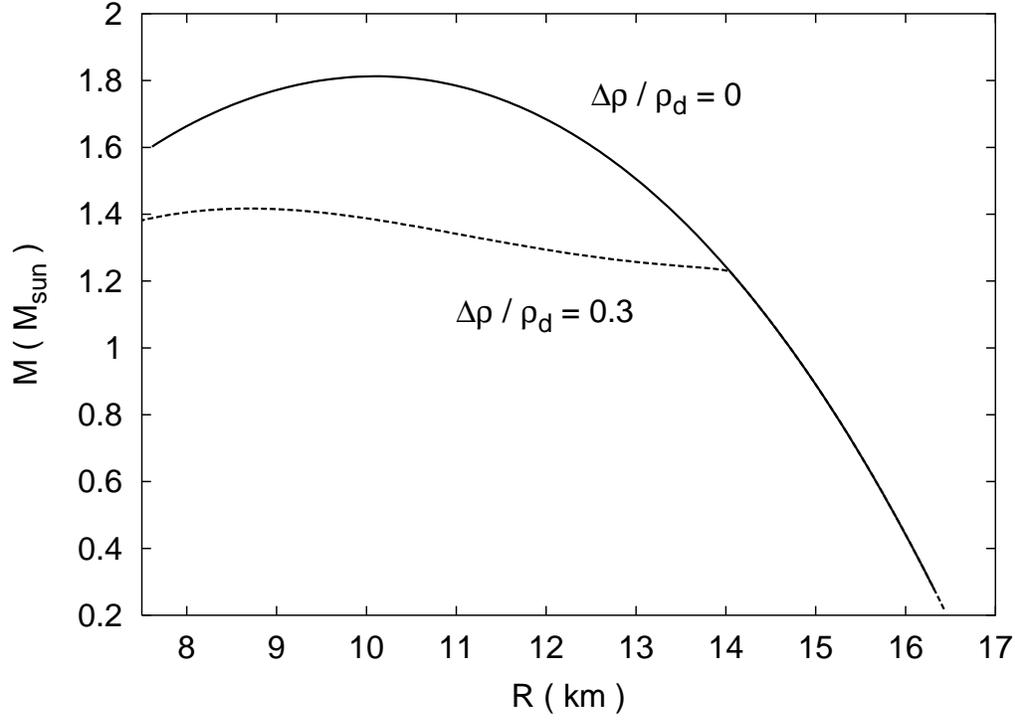,angle=-90,width=14cm}
\caption{{We compare the mass-radius curve of  stellar models without
discontinuity (continuous line) and with a high density 
discontinuity in the core (dashed line). Both curves refer to
the same low-density equation of state with $\Gamma = 2$, the only
difference being the presence of a discontinuity at $\rho_{\rm d} =
7\cdot 10^{14}~g/cm^3$ with $\Delta\rho / \rho_{\rm d} = 0.3,$ for the model
shown with a dashed line. As discussed in the text, 
when a discontinuity is present  the star is more compact and
the maximum mass is lower.
}}
\label{MRpoly}
\end{figure}
\end{center}

Before discussing the behaviour of the quasi normal mode frequencies, let
us shortly describe how the structure of a neutron star  changes if a phase
transition associated to a density discontinuity occurs in its inner core.
In Figure \ref{MRpoly} we plot the mass-radius relation for  a star
with and without discontinuity. The parameters 
are $\rho_{\rm d}=7\cdot 10^{14} ~g/cm^3$ and
$~\Delta\rho /\rho_{\rm d}=0.3$. The polytropic index
is $\Gamma = 2$ in both cases,  and we set
$K(1+\Delta\rho/\rho_{\rm d})^2=180$~km$^2$, in
order to have the same equation of state for the two models when
$\rho <\rho_{\rm d}.$
Figure \ref{MRpoly} shows that the maximum mass is lower
for the model with a density discontinuity; this is a general
behaviour, which is due to the softening of the equation of state
introduced by the discontinuity. As a rule of thumb, for a fixed mass, a
star with a discontinuity is more compact than one with the same mass and
no discontinuity at all; for instance,  from Figure  \ref{MRpoly}
we see that for $M = 1.4~M_\odot$ the star
with no phase transition has a radius of $R = 13.44$~km, while the one with
a density discontinuity has $R = 9.66$~km. These general properties continue to  hold 
also if  EOSs more realistic than eq. (\ref{eos}) are used, and if phase
transitions due to kaon/pion condensation or to the nucleation of quark
matter are considered \cite{heise}. Thus 
the EOS we use, though approximate, allows us to infer the 
mode properties that are known to depend much more on the star's macroscopic properties (mass,
radius, etc.), than on the specific microscopic interactions.

In our study, we shall consider stars with
a fixed mass $M = 1.4~M_\odot,$  because astronomical observations show
that NSs that have not spent a
significant part of their life accreting matter from a companion in a
binary system,  have masses in a narrow range around this fiducial
value \cite{TC99}. We shall choose different values of the polytropic
exponent $\Gamma=(1.67,~1.83,~2.00,~2.25)$, and assign the 
constant $K$ in such a way that models with the same $\Gamma$
have the same EOS for $\rho < \rho_{\rm d}~$, independently of the
particular  values of $\Delta\rho /\rho_{\rm d}$ and $\rho_{\rm d}$. Since
we require that all stars have the same mass, we will have to adjust the
central density accordingly. This means that different stars have a
different matter content in their core. Furthermore, we select models with
reasonable radii ($ 8.5~{\rm{km}} \leq R \leq 17.9~{\rm{km}}$) and,
since we want to study high density discontinuities, we fix their location
at $R_{\rm d} \leq 0.8~R.$  The amplitude of the
discontinuity is chosen to vary within $\Delta\rho /\rho_{\rm d}=(0.1-0.3)$;
larger values would not allow
stable stellar models with reasonable values of mass and radius.

In Tables \ref{tabres2}--\ref{tabres5} we summarize the
results of the numerical integration of the Eqs. (\ref{ldeq}) and
(\ref{zer})
for  these stellar models, as well as the equilibrium parameters. The
frequencies of the fundamental mode ($\nu_{\rm f}$), the first
pressure mode ($\nu_{\rm p}$), and
the discontinuity g--mode ($\nu_{\rm g}$) are tabulated for different
values of the parameters.
The models considered in Table \ref{tabres2} have the same value of
the polytropic exponent,   $\Gamma = 2$, and refer to three different
values of $\Delta\rho /\rho_{\rm d} = (~0.1, 0.2, 0.3~)$. 
In the remaining three Tables (\ref{tabres3}-\ref{tabres5}), we change the polytropic index
and for each value we consider a different  $\Delta\rho /\rho_{\rm d}.$
These choices allow to study the mode  frequencies for a wide range of stellar models.

The dependency of the mode frequencies on the variables relevant
to our analysis is better clarified in the following figures. In Figure
\ref{fpmrd}, we plot the frequency of the fundamental mode 
and of the first p--mode as a
function of $\rho_{\rm d},$ for  models with  $\Gamma = 2~$ (Table \ref{tabres2}),
and for the three selected values of $\Delta\rho
/\rho_{\rm d}$.
We choose  the $\Gamma = 2$ models because they are sufficiently
representative of the general behaviour.
The curves representing $\nu_{\rm f}$ and $\nu_{\rm p}$
for different $\Delta\rho /\rho_{\rm d}$ tend to a limiting value
(shown as a black dot in the Figure), which is the frequency of the
modes when there is no discontinuity inside the
star, i.e. when $\rho_{\rm d}$ exceeds the central density.
Both  $\nu_{\rm f}~$ and $\nu_{\rm p}~$ increase with
$\Delta\rho /\rho_{\rm d}$, and decrease with $\rho_{\rm d}.$ 
This behaviour is not surprising. Indeed,
from the data in Tables \ref{tabres2}--\ref{tabres5} we see that, having
fixed the mass of the NS, both the mean density and the compactness
increase with $\Delta\rho /\rho_{\rm d}$ and decrease with $\rho_{\rm d}$;
from the Newtonian theory of stellar perturbations it is known that 
the f--mode frequency is  proportional to  $\sqrt{M/R^3}~$,
and the first p--mode frequency scales with the
star's compactness $M/R$; thus, the  results shown in Figure 2 confirm this general behaviour, 
despite the structural changes produced by the density discontinuity. 
The dependency of $\nu_{\rm f}~$ and $\nu_{\rm p}~$  on 
the mean density and compactness of the star  has been studied in general relativity
for a large number of realistic EOSs;
empirical relations have been derived between the mode frequencies and the
parameters of the star, that  could be used
to infer the mass and the radius of neutron stars if a gravitational signal 
emitted by these pulsation modes  were detected \cite{nk1,nk2,nk3}.
These fits were obtained on the assumption that the density is continuous
inside the star, and it is interesting to see whether a discontinuity introduces deviations 
from the expected behaviour. For instance, in Figure 3 we plot, as a continuous line,
the linear fit which relates $\nu_{\rm f}$ and $\sqrt{M/R^3}~$
obtained by Andersson and Kokkotas (1998) (AK98-fit). 
In the same figure, we also  plot the values of $\nu_{\rm f}$ which we find 
for models with different values of $\Gamma$ and
$\Delta\rho /\rho_{\rm d}$. 
The frequencies of the models
with $\Delta\rho /\rho_{\rm d} = 0$ are shown as black dots.
We see that if a density discontinuity occurs in the 
inner core, $\nu_{\rm f}$  deviates from the
fit;  the fit  is expected to have an error less than $1\%$
\cite{nk3}, but the deviation introduced by a density discontinuity
is certainly larger.
Thus, the detection of an f--mode frequency would still allow to
estimate the mean density of the star, but with a larger uncertainty. 
Similar conclusions can be drawn by plotting the p--mode frequency as a function of $M/R$.

Let us now focus on the properties of the g--mode which is produced by the
density discontinuity. The  possible values of $\nu_{\rm g}$ cover a wide range,
$\nu_{\rm{g}}\approx [~0.5 - 1.4~]$~kHz (see Tables
\ref{tabres2}--\ref{tabres5}), and they are the highest
frequency g--modes that one can expect in neutron stars. Indeed, 
g--modes due to finite temperature effects have very low frequencies 
\cite{McD83,McD88}, while those due to composition gradients in 
neutron star cores
\cite{rg92,lai94} and to density discontinuities in the outer layers 
\cite{Finn87,McDermott90}  have, typically,  $\nu \le 200$~Hz. 
Thus, should a g--mode be detected, one could establish if it is due
to a discontinuity at supranuclear densities by looking at its
frequency. The observation of such a g--mode  would 
also  carry information on the parameters that characterize the
discontinuity. This is clear from Figure \ref{gmrd}, where we
plot $\nu_{\rm g}$ as a function of $\rho_{\rm d}$, for all the considered
models.  We see that the  g--mode frequencies belonging to the same 
density jump, $\Delta\rho /\rho_{\rm d},$ cluster in a definite region of frequency,
independently of the other parameters of the EOS
($\Gamma$ and $\rho_{\rm d}$), and this  is a  remarkable  feature.

Furthermore, for a fixed value of $\Delta\rho /\rho_{\rm d},$ $\nu_{\rm
g}$ exhibits a maximum as a function of the location of the discontinuity.
To better show this behaviour, in Figure \ref{gr} we explicitly plot
$\nu_{\rm g},~$ computed for two stars with the same $\Delta\rho/\rho_{\rm d}= 0.2$
and having $\Gamma = 1.83$ and $\Gamma =
2$, respectively,
as a function of the radius at which the discontinuity occurs,
$R_{\rm d}.$ The reason why there is such maximum is the
following. In a chemically homogeneous, zero temperature star with no
density discontinuity, the g--mode spectrum is degenerate at zero
frequency. When we consider stars with $R_{\rm d}~$ very close to center,
or when we put the discontinuity very close to the surface,
we are approaching the model with no discontinuity, and in both
cases $\nu_{\rm g}~$ tends to zero. Thus a maximum has to be expected at
some $R_{\rm d}~$, intermediate between these two extreme cases. For all
the models we studied the maximum g--mode frequency is found when $0.4
\leq R_{\rm d} /R \leq 0.6$.

As already shown by Finn (1987)  following a Newtonian analysis \cite{LL},
if the discontinuity is in the low-density region near the surface of the
star, the dispersion relation for the oscillation modes  can be written as
\begin{equation}
\omega^2_g \simeq l(l+1)\,\bar\rho\,\frac{\Delta\rho/\rho_{\rm d} }{1
+ \Delta\rho/\rho_{\rm d} }\,\frac{\Delta r}{R} \ ,
\label{llf}
\end{equation}
where $\bar\rho = M/R^3$ is the mean density of the star, $\Delta r = R-R_{\rm d},$
and $l$ is the order of the considered multipole. 
This relation is obtained by solving a stratified
two-fluid problem, taking the lower density fluid to be infinitely deep
and approximating the spherical symmetry of the star by plane symmetry,
assumptions that are reasonable if the density discontinuity is
near the surface of the star. Obviously, they are no longer satisfied when 
the discontinuity occurs at high density and $R_{\rm d} \approx 0.5~R$.
However, we find that the maximum g--mode frequency
for each assigned EOS satisfies a similar relation
\begin{equation}
\label{llf2}
   \omega^{max~2}_{g} \simeq l(l+1)\,{\mathcal C}^2\,\frac{\Delta r}{R} \ ,
\end{equation}
where 
\begin{equation}
{\mathcal C}^2 \equiv \bar{\rho}_{\rm i} \,\frac{\Delta\rho/\rho_{\rm d}}{1
          + {\Delta\rho}/\rho_{\rm d}} \ ,
\label{defC}
\end{equation}
which differs from the corresponding term in eq. (\ref{llf}) only because 
the mean density is replaced by the mean density of the ``internal'' core ($\rho > \rho_{\rm
d}$), $\bar{\rho}_{\rm i}.$
Considering all our stellar models, we find the following linear
fit 
\begin{equation}
\nu^{\rm max}_{\rm g} = \left[\,0.070
+ 0.905 \cdot \left( \frac{{\mathcal C}}{0.025~{\rm{km}}^{-1}}
\right)\,\right] \;\;{\rm kHz} \,
\label{Cfit}
\end{equation}
with an error smaller than $2\%$. For our models,
$0.016~{\rm{km}}^{-1}~\leq {\mathcal C} \leq 0.037~{\rm{km}}^{-1}$.
This linear behaviour has an interesting consequence. Suppose
that a g--mode with a given frequency $\nu_{\rm obs}$ could be
detected. Then, from Figure \ref{gmrd}
we would be able to constrain the amplitude of the density discontinuity
$\Delta\rho /\rho_{\rm d},$ and from  the fit 
(\ref{Cfit}) we could  set a lower limit on
${\mathcal C} $ and consequently  on
$\bar{\rho}_{\rm i}.$ All the EOSs  with
${\mathcal C} < {\mathcal C} ({\nu_{\rm{obs}}})$ would be ruled out
because they would have a g--mode with a maximum frequency lower than the
observed one.

\section{Excitation mechanisms and detectability}
We shall discuss three astrophysical processes that could be associated
to the excitation of discontinuity g--modes and to the emission of
gravitational waves: the coalescence  of a  NS-NS
binary system, the onset of a phase transition in the neutron star core,
and a pulsar glitch. For the last two cases, we shall estimate how much energy
should be deposited into  the modes in order the emitted signal to be
detectable by interferometric detectors very sensitive in the kHz
frequency region.

The tidal excitation of (non rotating) NS oscillations in binary
systems has been studied extensively in the past. In the case of 
core g--modes, it has been shown that the overlap integral between 
the displacement field and
the tidal force field is small \cite{rg94,holai,lai94}, and since the
energy  absorbed by a mode is proportional to the square of this integral,
they are basically ineffective. To ascertain 
whether  g--modes  due to density discontinuities could be more efficiently 
excited in a binary system, we adopt the formalism
introduced in previous works \cite{pap1,pap2}, where we computed the
gravitational perturbations of a neutron star induced  by an orbiting
companion by using a perturbative approach. We find that the typical resonance width
of a discontinuity g--mode ($\Delta\nu \approx 10^{-2}$~Hz)
is much smaller than that  of the f--mode ($\Delta\nu
\approx 100-200$~Hz).
Similarly to what happens for the g--modes arising from thermal and composition
gradients, the shift that the resonant excitation of a discontinuity g--mode produces
in the number of cycles of the emitted signal
is of the order of $10^{-2}$ cycles, too small to affect the detection significantly. 
For this reason,
tidal excitation of g--modes during the coalescence of NS
binary systems does not seem to be an efficient mechanism for the
emission of gravitational radiation, and we shall not consider it in
the following.

A second excitation mechanism is the onset of a phase transition.
According to current theories of stellar structure and evolution, the
matter composing a neutron star may undergo a phase transition to quark
matter if, at some point of the star life,  a sufficiently high density is
reached in its core.  The critical density required to produce such a
transition can be reached soon after the core-collapse which gives birth to the
neutron star, or at a later stage of the evolution, due to accretion
\cite{chengdai} and/or spin-down \cite{maxie}. The onset of a phase
transition is accompanied by a sudden reduction of the stellar radius and
the difference in binding energy between the initial and final
configuration is expected to be radiated away. The binding energy released
during this late {\it mini-collapse} is of the same order as that emitted
during neutron star formation following a gravitational core--collapse, i.e. $
\approx 2 \times 10^{53} {\rm erg} = 0.1 M_\odot c^2$ \cite{AFO86,datta}.
How much of this energy reservoir can be redistributed in  the different
modes of oscillation of the star and radiated in gravitational wave is,
however,  an open issue. Therefore,  the only thing we can reasonably
estimate is which fraction of this energy should go into 
the quasi normal modes in
order the signal emitted at the corresponding frequencies to  be detectable.

The quasi normal modes of oscillation of a star may also be 
excited as a consequence of a glitch. 
Glitches are sudden changes in the rotation frequency of the
neutron star crust which superimpose to the usual gradual spin-down under
magnetic torque. They are observed in many pulsars and are thought to be
related to quakes occurring in the solid structures such as the crust, the
superfluid vortices and, perhaps, the lattice of quark matter in the
stellar core \cite{gl1,gl2,gl3}. The observed glitches are very small,
with a typical fractional spin variation of about $\Delta\Omega /\Omega
\approx 10^{-6} - 10^{-8}$, which would be associated to an
energy release of the order of
  $~\Delta E \approx I\Omega\Delta\Omega$, where $I$ is the
moment of inertia of the star. For the glitches observed in the Crab and Vela pulsars
this simple estimate gives
$~\Delta E \simeq 2\cdot 10^{-13}~M_\odot c^2$ and $~\Delta E \simeq 3\cdot
10^{-12}~M_\odot c^2$, respectively. As before, we shall assume that a
fraction of this energy is redistributed into the quasi normal modes of
oscillation of the perturbed star, and we shall evaluate how much is
needed to excite a mode to a detectable level.

\subsection{Detectability}
The frequency of the discontinuity g--modes we have studied ranges between
$\nu_{\rm g} \simeq [0.5-1.4]$~kHz, while  $\nu_{\rm f} \simeq
[1.3-3.2]$~kHz and for the first p-mode
$\nu_{\rm p} \simeq [2.5-7.3]$~kHz. In order to detect a signal emitted as a
consequence of the excitation of these modes a very sensitive high
frequency detector would be needed. In the following we shall consider the
gravitational laser interferometric detector EURO, that has recently  been
proposed in a preliminary assessment study and for which the sensitivity
has been estimated by Sathyaprakash and Schutz
(http://www.astro.cf.ac.uk/geo/euro). EURO should be
extremely sensitive in the kHz region, which is crucial to
infer informations about neutron stars interior from gravitational wave
data, and it has been envisaged in  two experimental configurations: one
for which the noise curve  in the frequency region $10-10^4~Hz$ can be
written as
\begin{equation}
\label{euroshot} S_n (\nu ) = 10^{-50}\,\left[\frac{3.6~10^9}{\nu^4} +
\frac{1.3~10^5}{\nu^2} +
             1.3~10^{-3}\,\nu_k\,
             \left(1+\frac{\nu^2}{\nu_k^2}\right)\right] {\mathrm Hz}^{-1}\ ,
\end{equation}
where $\nu_k = 10^3~Hz$; the second  configuration is the more ambitious Xylophone detector,
for which the shot noise, i.e. the third term within brackets in (\ref{euroshot}),
should be eliminated by running several narrow-banded interferometers.
We will refer to these two possible configurations as EURO and EURO-Xylo,
respectively. We shall assume that the gravitation signal from a star
pulsating in a quasi-normal mode is a damped sinusoid of the form
\cite{echeve}
\begin{equation}
\label{eche}
 h (t) = {\cal{A}} e^{(t_{\rm arr} -t)/\tau} \sin
      \left[2\pi \nu \left(t - t_{\rm arr}\right)\right]\ ,
\end{equation}
where $t_{\rm arr}$ is the arrival time and $\tau$ the damping time of the
oscillation. The amplitude ${\cal A}$ can be expressed in terms of the
total energy $\Delta E_\odot = \Delta E /M_\odot c^2$ deposited into the
mode, and of the distance $D$ of the source by \cite{nk3}
\begin{equation}
\label{amp}
 {\cal A} \simeq 7.6~10^{-23}\,\sqrt{\frac{\Delta E_\odot}{10^{-10}}\,
            \frac{1~{\rm s}}{\tau}}\,
   \left(\frac{1~{\rm{kpc}}}{D}\right)\,\left(\frac{10^3~{\rm{Hz}}}{\nu}\right) \ .
\end{equation}
Given a detector with noise spectrum $S_n(\nu)$, and assuming that the
matched filtering technique is used to extract the signal,
the signal to noise ratio (SNR), can be written as
\begin{equation}
\label{SN}
 SNR = \left[{\cal F}\,\frac{{\cal{A}}^2
\tau}{2S_n}\right]^{1/2} \ ,
\end{equation}
where $~{\cal F} = 4Q^2 / ( 1+4Q^2 )~$ is a form factor, and
$Q = \pi \nu \tau$ is the quality factor of the oscillation.

It should be noted that
in the case of discontinuity g--modes, since the gravitational damping time
ranges between $\tau \approx ( 10^7 - 10^{12} )~s$, the oscillations
will be damped by other dominant dissipative mechanisms rather than by the
emission of gravitational waves. In newly born neutron stars (or hybrid
stars with a quark core) the dominant damping mechanism is neutrino
viscosity, with a typical timescale of $10 - 10^2$~s
\cite{vis1,vis2,vis3}. 
In the glitches scenario (or other similar excitations occurring in late
stages of the evolution of neutron stars) the neutron star is cold and
neutrinos are not coupled to the matter. The dissipative mechanisms are
then associated to  the kinematical properties of matter. In this case,
bulk, shear viscosity and boundary layer dissipation are more efficient
than gravitational damping. In any case, the typical dissipative
time-scales are thought to be much larger than the neutrino damping 
time-scale in
proto-neutron(quark) stars \cite{Bild00}, but a word of caution is needed.
It has been found that quark-matter viscosity is much larger than that of
normal neutron star matter \cite{madsen}, and therefore it would quickly
damp the oscillations in the dense quark core ($\tau\sim 10^{-2}$ s).
However, these results are still under debate, and it  remains to be
carefully analyzed how the outer shell would react to the strong damping
in the inner region, and whether or not boundary effects would change the
picture.
Note however that, for the range of frequencies
considered in this paper, the quality factor $Q \gg 1$ as long as $\tau$
is greater that, say, $10^{-2}~s$; thus, the form factor in Eq. (\ref{SN})
can be taken equal to one, and the signal to noise ratio happens to be
quite insensitive to the damping time.

In the following, we shall estimate the amount of energy that should go
into a quasi normal mode to produce a gravitational signal 
detectable with a signal to noise ratio $SNR=3$, either by  EURO or by
 EURO-Xylo detector. For
g--mode pulsations, we consider the two limiting frequencies
$\nu_1 = 0.5$~kHz and $\nu_2 = 1.4$~kHz, being the g--mode
frequency of all our models included in this range.  The  excitation of a  core
g--mode at $0.1$~kHz \cite{rg92,lai94} is also computed for
comparison.
For the f-- and p--modes, our estimates are done for two 
frequencies, $\nu_{\rm f} = 2.3$~kHz and $\nu_{\rm p} = 5.5$~kHz,
which are obtained  as an average over the corresponding 
range of variation  for all our models.
The gravitational damping time of the fundamental and p--mode
are smaller than the other dissipative time scales;
thus, in this case the emission of gravitational waves is the dominant
dissipative mechanism  and both $\tau_{\rm f}$ and $\tau_{\rm p}$ have
been used explicitly  in our estimates.

For the two different excitation scenarios, mini-collapse and
glitches, we shall consider sources located at distances $D=15$~Mpc and
$D=10$~kpc, respectively. Note that  the Virgo cluster, with its some
$2~10^3$ members, is at $D \approx 15~Mpc$ and that the Vela and Crab pulsars,
that are known to produce glitches, are at distances of $D=0.5~kpc$ and
$D=2~kpc$, respectively.

Our estimates are given in Table (\ref{Eonset}), and show
that the high frequency detectors would be a very interesting
instrument to probe, through
gravitational signals, the nature of neutron stars interiors. Note
that, since for the EURO configuration the sensitivity at high
frequency is strongly limited by the shot noise, a p--mode 
at $\nu_{\rm p} = 5.5$~kHz could be detected only with the
advanced EURO-Xylo configuration.  If one focuses on the
fundamental mode  and the discontinuity  g--mode, the
minimal energies required to detect a signal with $SNR=3$ with the
EURO detector  range within
$1.6\cdot 10^{-13} \le \Delta E / M_\odot c^2 \le 3.0\cdot 10^{-5}$, 
depending on the source distance and on the frequency of the
signal. With the EURO-Xylo configuration, this range  becomes
$4.1\cdot 10^{-14} \le \Delta E / M_\odot c^2 \le 3.4\cdot 10^{-7}$.
As mentioned  before, the amount of binding energy which is expected to be
released in a mini-collapse is approximately the same as that emitted
during a core-collapse Supernova. In the Supernova case, it has been suggested that
the fraction  of this energy  radiated in gravitational waves  is in the range
 $[10^{-8}-10^{-6}]~M_\odot c^2$ \cite{coll1,coll3,coll4}; if we assume
that a comparable amount of energy goes in
gravitational waves also in the case of a mini-collapse, the chance
for a quasi normal mode signal to be detectable from a source
closer than 15~Mpc is promising. The situation is clearly even more 
interesting for galactic sources, since
the amount of energy required to excite a mode at a detectable level
is typically a small fraction of the total energy reservoir. 

\section{Conclusions}

According to the relativistic  theory of stellar perturbations,  a star
perturbed by any external or internal process  emits gravitational waves
at  the characteristic frequencies of its quasi normal modes. It has been
shown that both the mass and the radius of a pulsating cold neutron star
could be inferred, if the quasi normal modes were identified by a spectral
analysis in the detected wave \cite{nk2}. More recently,  the possibility
of tracing the presence of a superfluid core through the detection of
superfluid modes has also been investigated \cite{nilssup}. As described
in the introduction, g-modes are also good markers of the internal
composition of a neutron star: low frequency g-modes indicate a non
homogeneous composition in the outer layers or in the core of the star,
or a thermal profile, whereas the high frequency g-modes which we study in this paper, 
are  associated to phase transitions occurring at supranuclear densities 
(Sotani {\it et al.}, 2001).
Although the analysis has been restricted to simplified polytropic EOSs,
our results can be generalized to realistic equations of state, since
it is known that the oscillation frequencies depend more on global
properties, like mass and radius,  
than on the specific form of the microscopical interactions. 
We have analyzed  models of stars with a mass
of 1.4 M$_\odot$, since astrophysical observations show that most of the
well measured masses of neutron stars in binary systems cluster in
a narrow range around this value. 

We find that high frequency discontinuity g--modes
exhibit several interesting features:
i) the linear relation known to exist 
between the frequency of the fundamental mode 
and the square root of the average density,  still holds for
neutron stars with density discontinuities. Thus, a measure of the
f-mode frequency would be useful to constrain the neutron star
radius, provided the mass is measured independently,
even when a phase transition takes place in its core.
However, the presence of the
discontinuity introduces a larger error on the NS radius determination.
ii) Should density discontinuities actually exist, the associated g--mode
will have a frequency of about 1 kHz, somewhat lower than that of the
f-mode, and therefore clearly distinguishable. It may be reminded that
thermal or chemical g-modes appear at significantly lower frequencies
($<200$ Hz). iii) Since the g--mode frequency strongly depends on
the amplitude of the discontinuity, it could be used to infer
the value of $\Delta\rho /\rho_{\rm d}$. In addition, since the maximum of 
$\nu_{\rm g},$ considered as a function of the discontinuity radius $R_{\rm d}$,
is related to $\Delta\rho /\rho_{\rm d}$ and to the average density,  $\bar{\rho}_{\rm i},$ 
of the core internal to $R_{\rm d}$, (see eqs. (\ref{defC}) and (\ref{Cfit})) 
it would be possible to set a lower bound
on $\bar{\rho}_{\rm i},$  and to rule out all EOSs for which the combination
$~{\mathcal C}^2 \equiv \bar{\rho}_{\rm i} \,\frac{{\Delta\rho}/{\rho_{\rm d}}}{1
          + {\Delta\rho}/{\rho_{\rm d}}}~$
is smaller than the observed one.

In the spirit of previous works \cite{nilssup}, which discussed the
observability of oscillation modes produced by the existence of superfluid
components in the interior of a neutron star, 
we have analyzed some astrophysical processes in which
high frequency modes  could be excited,
and we have evaluated the amount of
energy that the modes should convey to produce a gravitational signal 
detectable with a SNR=3  by the high frequency
gravitational detectors under study.
We find that the required  energy does 
not conflict with the  estimates of the 
energy  expected to be released in gravitational waves at the onset 
of a phase transition, or during a glitch.
In particular, the detector  EURO, especially in the configuration known as
EURO-Xylophone, would have a  significant chance to reveal the gravitational
signal associated to the excitation of the fundamental mode, of the first
pressure mode and of the discontinuity g--mode following a galactic glitch or
a stellar core-collapse or a mini-collapse induced by the onset of a
phase transition to quark matter within 10-15 Mpc.  
By cross-correlating the information carried by these
mode frequencies, and in particular by the high frequency discontinuity g--modes,
we would acquire  extremely valuable indications 
on the properties of nuclear matter at supranuclear density.

\begin{center}
\bf{Acknowledgements}
\end{center}

We would like to thank O. Benhar for useful discussions. We thank
B.S. Sathyaprakash for providing us the power spectral density of the
EURO detector.

This work has been supported by the EU Programme 'Improving the Human
Research Potential and the Socio-Economic Knowledge Base' (Research
Training Network Contract HPRN-CT-2000-00137). JAP is supported by the
Marie Curie Fellowship No. HPMF-CT-2001-01217.


\clearpage

\begin{table}
\begin{center}
\begin{tabular}{||p{0.01cm}*{8}{c|}|}
\hline
& $R~(km)$ & $\rho_{\rm c}~(g/cm^3) $ & $\sqrt{\bar\rho}~(km^{-1}) $ &
$\rho_{\rm{d}}~(g/cm^3)$  & $\Delta\rho /\rho_{\rm d}$ &
$\nu_{\rm{f}}~(kHz)$ & $\nu_{\rm p}~(kHz)$& $\nu_{\rm g}~(kHz)$ \\
\hline
& $13.44$ &$0.92~10^{15}$ & $0.0292$ & $--$ & $0.0$ & $1.666$ &
$4.045$ & $--$ \\
\hline
\hline
& $12.04$ & $1.39~10^{15}$& $0.0344$ & $3~10^{14}$  & $0.1$ & $1.998$ &
$4.637$ & $0.504$ \\
\hline
& $12.22$ & $1.36~10^{15}$ & $0.0337$ & $4~10^{14}$  & $0.1$ & $1.962$ &
$4.548$ &$0.567$ \\
\hline
& $12.42$ & $1.32~10^{15}$& $0.0329$ & $5~10^{14}$  & $0.1$ & $1.915$ &
$4.459$ & $0.613$ \\
\hline
& $12.65$ & $1.27~10^{15}$ & $0.0319$ & $6~10^{14}$  & $0.1$ & $1.857$ &
$4.365$ & $0.644$ \\
\hline
& $12.92$ & $1.19~10^{15}$ & $0.0310$ & $7~10^{14}$  & $0.1$ & $1.792$ &
$4.260$ & $0.659$ \\
\hline
& $13.21$ & $1.11~10^{15}$ & $0.0300$ & $8~10^{14}$  & $0.1$ & $1.723$ &
$4.146$ & $0.658$ \\
\hline
& $13.43$ & $1.02~10^{15}$ & $0.0292$ & $9~10^{14}$  & $0.1$ & $1.670$ &
$4.052$ & $0.641$ \\
\hline
\hline
& $10.71$ & $2.17~10^{15}$ & $0.0410$ & $4~10^{14}$  & $0.2$ & $2.408$ &
$5.325$ & $0.840$ \\
\hline
& $10.99$ & $2.07~10^{15}$ & $0.0395$ & $5~10^{14}$  & $0.2$ & $2.330$ &
$5.155$ & $0.912$ \\
\hline
& $11.35$ & $1.95~10^{15}$ & $0.0376$ & $6~10^{14}$  & $0.2$ & $2.226$ &
$4.970$ & $0.961$ \\
\hline
& $11.83$ & $1.77~10^{15}$ & $0.0354$ & $7~10^{14}$  & $0.2$ & $2.088$ &
$4.750$ & $0.987$ \\
\hline
& $12.50$ & $1.51~10^{15}$ & $0.0325$ & $8~10^{14}$  & $0.2$ & $1.901$ &
$4.452$ & $0.979$ \\
\hline
& $13.38$ & $1.15~10^{15}$ & $0.0294$ & $9~10^{14}$  & $0.2$ & $1.680$ &
$4.072$ & $0.906$ \\
\hline
\hline
& $8.68$ & $4.44~10^{15}$ & $0.0562$ & $5~10^{14}$  & $0.3$ & $3.216$ &
$6.683$ & $1.211$ \\
\hline
& $9.13$ & $3.95~10^{15}$ & $0.0521$ & $6~10^{14}$  & $0.3$ & $3.039$ &
$6.325$ & $1.281$ \\
\hline
& $9.66$ & $3.46~10^{15}$ & $0.0479$ & $7~10^{14}$  & $0.3$ & $2.831$ &
$5.953$ & $1.326$ \\
\hline
& $10.41$ & $2.88~10^{15}$ & $0.0428$ & $8~10^{14}$  & $0.3$ & $2.553$ &
$5.506$ & $1.339$ \\
\hline
& $12.15$ & $1.87~10^{15}$ & $0.0340$ & $9~10^{14}$  & $0.3$ & $2.002$ &
$4.632$ & $1.251$ \\
\hline
\end{tabular}
\end{center}
\caption{The  frequencies of the fundamental mode, of the first p--mode
and of the discontinuity g--mode are shown for a set of stellar models
with $M=1.4~M_\odot$ and $\Gamma = 2$.
For each star, the equilibrium  parameters are also tabulated.
The polytropic coefficient $K$ is 
$K ( 1 +\Delta\rho /\rho_{\rm d} )^2 = 180~km^2 $, so that all  stars
have the same low-density equation of state. The first entry
corresponds to a model with no density discontinuity.} \label{tabres2}
\end{table}

\clearpage

\begin{table}
\begin{center}
\begin{tabular}{||p{0.01cm}*{8}{c|}|}
\hline
& $R~(km)$ & $\rho_{\rm c}~(g/cm^3) $ & $\sqrt{\bar\rho}~(km^{-1}) $ &
$\rho_{\rm{d}}~(g/cm^3)$  & $\Delta\rho /\rho_{\rm d}$ &
$\nu_{\rm{f}}~(kHz)$ & $\nu_{\rm p}~(kHz)$& $\nu_{\rm g}~(kHz)$ \\
\hline
& $17.96$ &$0.72~10^{15}$ & $0.0189$ & $--$ & $0.0$ & $1.286$ &
$2.543$ & $--$ \\
\hline
\hline
& $13.38$ & $2.14~10^{15}$& $0.0294$ & $4.0~10^{14}$  & $0.1$ & $2.035$ &
$3.747$ & $0.641$ \\
\hline
& $14.05$ & $1.86~10^{15}$& $0.0273$ & $4.5~10^{14}$  & $0.1$ & $1.895$ &
$3.528$ & $0.648$ \\
\hline
& $14.75$ & $1.61~10^{15}$& $0.0254$ & $5.0~10^{14}$  & $0.1$ & $1.764$ &
$3.321$ & $0.650$ \\
\hline
& $15.51$ & $1.38~10^{15}$& $0.0235$ & $5.5~10^{14}$  & $0.1$ & $1.633$ &
$3.311$ & $0.646$ \\
\hline
& $16.39$ & $1.15~10^{15}$& $0.0217$ & $6.0~10^{14}$  & $0.1$ & $1.495$ &
$2.892$ & $0.635$ \\
\hline
& $17.32$ & $0.94~10^{15}$& $0.0199$ & $6.5~10^{14}$  & $0.1$ & $1.367$ &
$2.680$ & $0.614$ \\
\hline
& $17.90$ & $0.81~10^{15}$& $0.0190$ & $7.0~10^{14}$  & $0.1$ & $1.293$ &
$2.555$ & $0.593$ \\
\hline
\end{tabular}
\end{center}
\caption{ The equilibrium parameter and the quasi normal mode frequencies are shown as in
Table 1 for stellar models with $M=1.4~M_\odot$ and $\Gamma = 1.67$. The 
polytropic coefficient $K$ is chosen in such a way that the 
star with no discontinuity ($\Delta\rho /\rho_{\rm d} = 0$ ) and those with
$\Delta\rho /\rho_{\rm d} = 0.1$ have the same low-density equation 
of state.} 
\label{tabres3}
\end{table}

\begin{table}
\begin{center}
\begin{tabular}{||p{0.01cm}*{8}{c|}|}
\hline
& $R~(km)$ & $\rho_{\rm c}~(g/cm^3) $ & $\sqrt{\bar\rho}~(km^{-1}) $ &
$\rho_{\rm{d}}~(g/cm^3)$  & $\Delta\rho /\rho_{\rm d}$ &
$\nu_{\rm{f}}~(kHz)$ & $\nu_{\rm p}~(kHz)$& $\nu_{\rm g}~(kHz)$ \\
\hline
& $15.17$ &$0.82~10^{15}$ & $0.0243$ & $--$ & $0.0$ & $1.490$ &
$3.308$ & $--$ \\
\hline
\hline
& $10.22$ & $3.40~10^{15}$& $0.0440$ & $4.0~10^{14}$  & $0.2$ & $2.736$ &
$5.361$ & $0.932$ \\
\hline
& $10.51$ & $3.16~10^{15}$& $0.0422$ & $4.5~10^{14}$  & $0.2$ & $2.639$ &
$5.185$ & $0.961$ \\
\hline
& $10.82$ & $2.94~10^{15}$& $0.0404$ & $5.0~10^{14}$  & $0.2$ & $2.542$ &
$5.016$ & $0.985$ \\
\hline
& $11.17$ & $2.72~10^{15}$& $0.0385$ & $5.5~10^{14}$  & $0.2$ & $2.434$ &
$4.839$ & $1.003$ \\
\hline
& $11.58$ & $2.48~10^{15}$& $0.0365$ & $6.0~10^{14}$  & $0.2$ & $2.314$ &
$4.648$ & $1.014$ \\
\hline
& $12.07$ & $2.22~10^{15}$& $0.0343$ & $6.5~10^{14}$  & $0.2$ & $2.177$ &
$4.434$ & $1.016$ \\
\hline
& $12.71$ & $1.92~10^{15}$& $0.0317$ & $7.0~10^{14}$  & $0.2$ & $2.011$ &
$4.174$ & $1.007$ \\
\hline
& $13.66$ & $1.53~10^{15}$& $0.0285$ & $7.5~10^{14}$  & $0.2$ & $1.789$ &
$3.817$ & $0.972$ \\
\hline
& $15.08$ & $1.03~10^{15}$& $0.0246$ & $8.0~10^{14}$  & $0.2$ & $1.507$ &
$3.338$ & $0.874$ \\
\hline
\end{tabular}
\end{center}
\caption{ The equilibrium parameter and the quasi normal mode frequencies are shown 
for stellar models with $M=1.4~M_\odot$ and $\Gamma = 1.83$. The 
first entry is a star with no discontinuity, the other stars have a 
discontinuity of amplitude $\Delta\rho /\rho_{\rm d} = 0.2,$
and  the polytropic coefficient $K$ is chosen  as in Table 2.}
\label{tabres4}
\end{table}

\clearpage

\begin{table}
\begin{center}
\begin{tabular}{||p{0.01cm}*{8}{c|}|}
\hline
& $R~(km)$ & $\rho_{\rm c}~(g/cm^3) $ & $\sqrt{\bar\rho}~(km^{-1}) $ &
$\rho_{\rm{d}}~(g/cm^3)$  & $\Delta\rho /\rho_{\rm d}$ &
$\nu_{\rm{f}}~(kHz)$ & $\nu_{\rm p}~(kHz)$& $\nu_{\rm g}~(kHz)$ \\
\hline
& $11.32$ &$1.22~10^{15}$ & $0.0378$ & $--$ & $0.0$ & $2.005$ &
$5.361$ & $--$ \\
\hline
\hline
& $8.53$ & $3.44~10^{15}$& $0.0577$ & $6.0~10^{14}$  & $0.3$ & $3.095$ &
$7.329$ & $1.171$ \\
\hline
& $8.71$ & $3.34~10^{15}$& $0.0560$ & $7.0~10^{14}$  & $0.3$ & $3.031$ &
$7.145$ & $1.253$ \\
\hline
& $8.92$ & $3.21~10^{15}$& $0.0540$ & $8.0~10^{14}$  & $0.3$ & $2.945$ &
$6.949$ & $1.317$ \\
\hline
& $9.20$ & $3.03~10^{15}$& $0.0516$ & $9.0~10^{14}$  & $0.3$ & $2.831$ &
$6.734$ & $1.362$ \\
\hline
& $9.58$ & $2.77~10^{15}$& $0.0485$ & $1.0~10^{15}$  & $0.3$ & $2.673$ &
$6.473$ & $1.395$ \\
\hline
& $10.14$ & $2.41~10^{15}$& $0.0445$ & $1.1~10^{15}$  & $0.3$ & $2.445$ &
$6.116$ & $1.368$ \\
\hline
& $10.60$ & $2.12~10^{15}$& $0.0417$ & $1.15~10^{15}$  & $0.3$ & $2.268$ &
$5.829$ & $1.326$ \\
\hline
& $11.26$ & $1.66~10^{15}$& $0.0381$ & $1.2~10^{15}$  & $0.3$ & $2.027$ &
$5.404$ & $1.208$ \\
\hline
\end{tabular}
\end{center}
\caption{ The equilibrium parameter and the quasi normal mode frequencies are shown as in
previous Tables, for stellar models with $M=1.4~M_\odot,$  $\Gamma = 2.25$ 
and  $\Delta\rho /\rho_{\rm d} = 0.3.$}
\label{tabres5}
\end{table}

\begin{table}
\begin{center}
\begin{tabular}{||p{0.01cm}*{6}{c|}|}
\hline
& mode & $\nu_{\rm{mode}}$ &
\multicolumn{2}{c}{D = 15 Mpc } & \multicolumn{2}{c}{D = 10 kpc }
\\
& \multicolumn{2}{c}{ } &  $\Delta E $ (Euro) & $\Delta E $ (Euro-Xylo) &
$\Delta E $ (Euro) & $\Delta E $ (Euro-Xylo)
\\
\hline
\hline
& g$_{\rm{core}}$ & $0.1$~kHz & $3.5\cdot 10^{-7}~M_\odot c^2$ &
$3.4\cdot 10^{-7}~M_\odot c^2$
& $1.6\cdot 10^{-13}~M_\odot c^2$ & $1.5\cdot 10^{-13}~M_\odot c^2$
\\
\hline
& g$_{\rm{disc}}$ & $0.5$~kHz & $3.9\cdot 10^{-7}~M_\odot c^2$ &
$1.0\cdot 10^{-7}~M_\odot c^2$
& $1.7\cdot 10^{-13}~M_\odot c^2$ & $4.5\cdot 10^{-14}~M_\odot c^2$
\\
&  & $1.4$~kHz & $5.4\cdot 10^{-6}~M_\odot c^2$ &
$9.2\cdot 10^{-8}~M_\odot c^2$
& $2.4\cdot 10^{-12}~M_\odot c^2$ & $4.1\cdot 10^{-14}~M_\odot c^2$
\\
\hline
& f & $2.3$~kHz & $3.0\cdot 10^{-5}~M_\odot c^2$ &
$9.2\cdot 10^{-8}~M_\odot c^2$
& $1.4\cdot 10^{-11}~M_\odot c^2$ & $4.1\cdot 10^{-14}~M_\odot c^2$
\\
\hline
& p$_1$ & $5.5$~kHz & $8.7\cdot 10^{-4}~M_\odot c^2$ &
$9.1\cdot 10^{-8}~M_\odot c^2$
& $3.8\cdot 10^{-10}~M_\odot c^2$ & $4.0\cdot 10^{-14}~M_\odot c^2$
\\
\hline
\end{tabular}
\end{center}

\caption{\small{The amount of energy that should  be radiated in gravitational
waves at the frequency of a given mode, in order 
for the signal to be detectable with a $SNR=3$ by the EURO detector
configurations. The source distance is taken to be $D=15~Mpc$ 
and $D=10~kpc$ (see discussion in the text).}}
\label{Eonset}
\end{table}

\clearpage

\begin{center}
\begin{figure}
\psfig{figure=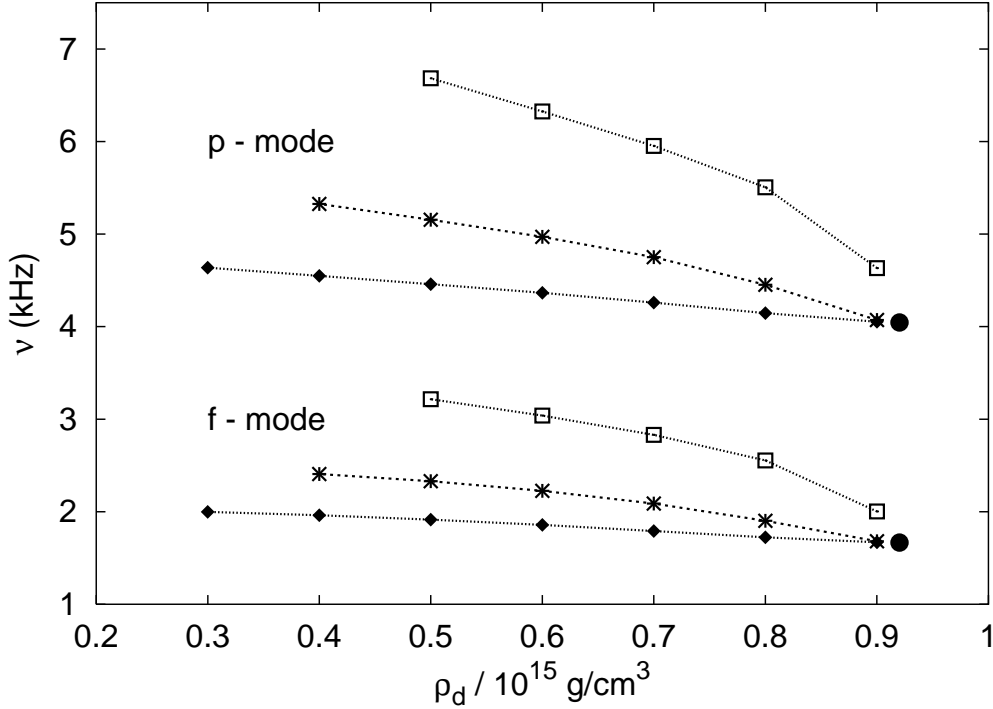,angle=-90,width=14cm}
\caption{The f--mode and the first  p--mode frequencies are plotted as a
function of the density at the discontinuity, $\rho_{\rm d}$, for the 
models of stars with mass $M=1.4~M_\odot$ and $\Gamma = 2$  given in
Table \ref{tabres2}.
The different curves refer to assigned values of  $\Delta\rho /\rho_{\rm d} = (~0.1,
0.2, 0.3~)$, increasing from bottom to top. The variation
of the mode frequency with the discontinuity parameters depends on the
associated variation of the mean density and the compactness of the star (see text).
All  mode frequencies tend to the same limiting value (black dot) as
$\rho_{\rm d}$ approaches the central density, which is the
frequency of the mode when there is  no discontinuity.
} \label{fpmrd}
\end{figure}
\end{center}

\begin{center}
\begin{figure}
\psfig{figure=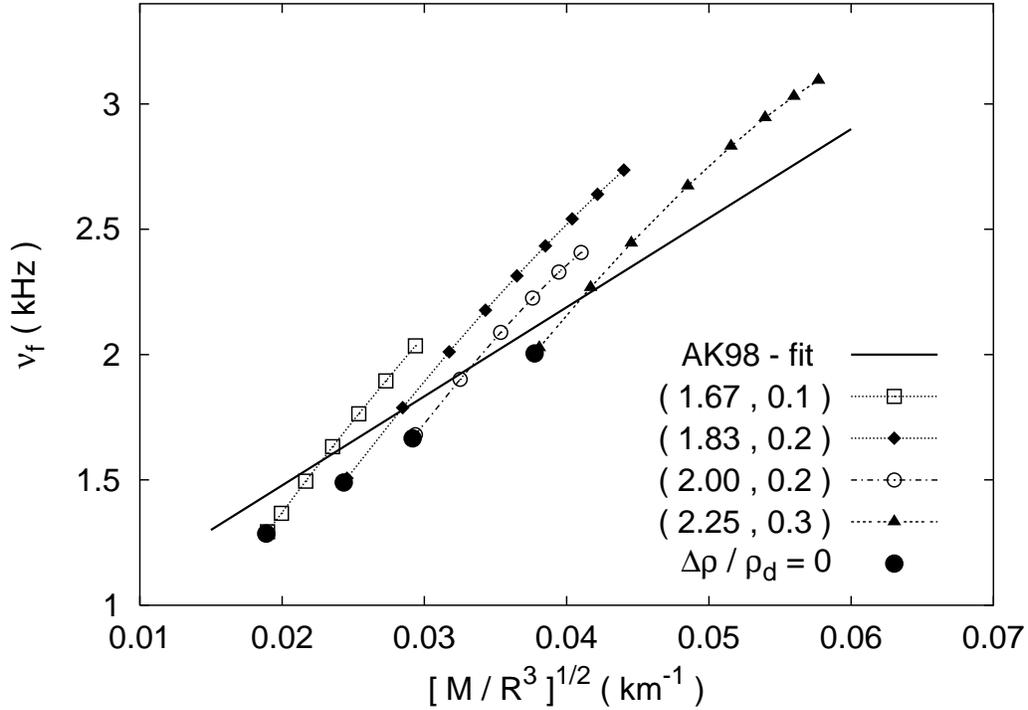 ,angle=-90,width=14cm}
\caption{We show the f--mode frequency  as a function of the mean density of
the NS for a sample of our models, labelled with
($\Gamma$, $\Delta\rho /\rho_{\rm d}$). The fit by Andersson and Kokkotas
(1998), obtained for stars without density discontinuities, is shown as a continuous
line (AK98-fit), while the black dots are the
f--mode frequency of our models with $\Delta\rho /\rho_{\rm d} = 0$. The
Figure shows that, for each assigned value of  $\Delta\rho /\rho_{\rm d}$, 
the f--mode frequency
is still a linear function of $\sqrt{M/R^3}$, as in the case of stars with
no discontinuity, but with a different slope.}
\label{fmrho}
\end{figure}
\end{center}

\begin{center}
\begin{figure}
\psfig{figure=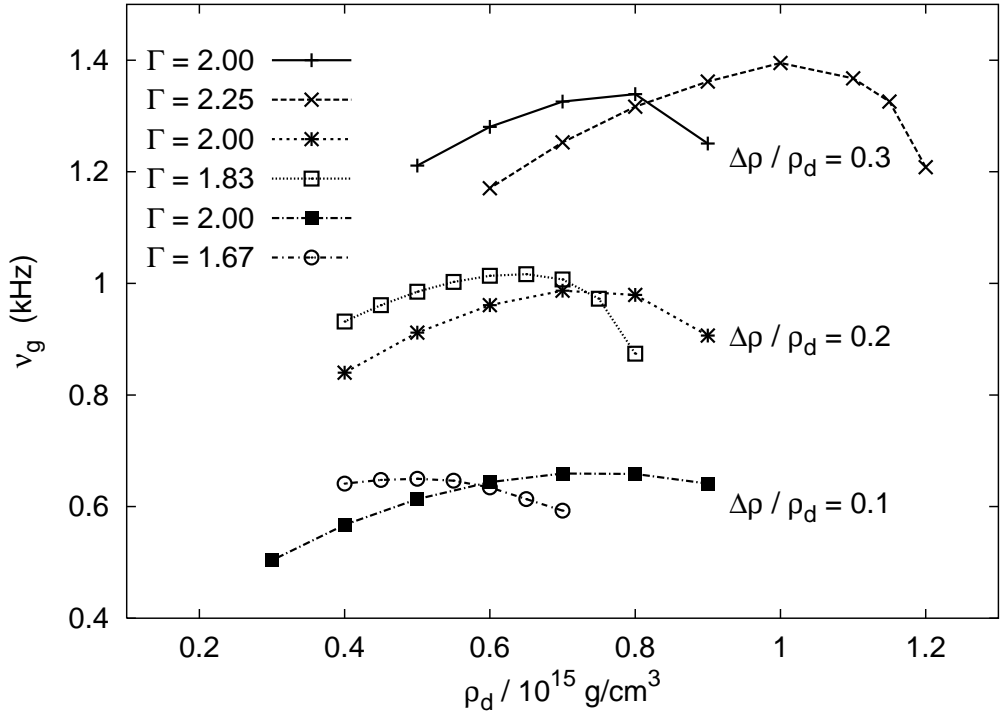,angle=-90,width=14cm}
\caption{The g--mode frequency of the stellar models given in 
Tables \ref{tabres2}--\ref{tabres5}, is shown as a function of $\rho_{\rm
d}$. It is remarkable that $\nu_{\rm g}$ depends much more on the amplitude of
the density discontinuity $\Delta\rho /\rho_{\rm d}$, than on the
other parameters of the EOS. Should a g--mode ever be detected, this feature would
allow to infer the value of the density jump.}
\label{gmrd}
\end{figure}
\end{center}

\begin{center}
\begin{figure}
\psfig{figure=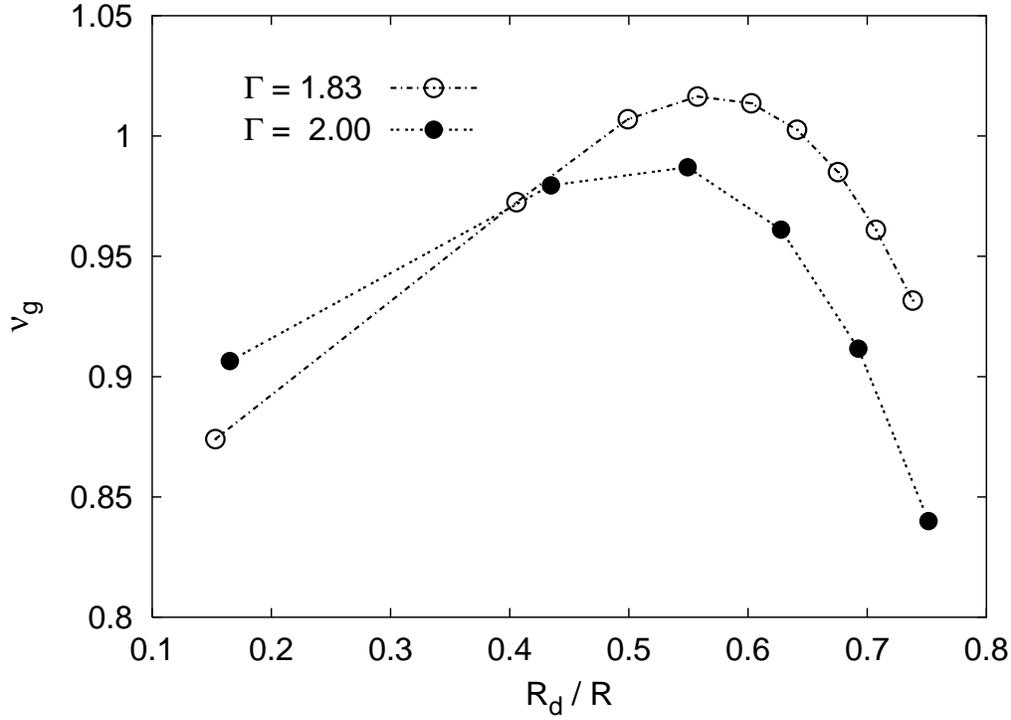,angle=-90,width=14cm}
\caption{ The discontinuity g--mode frequency is plotted as a function of
the normalized radius of the internal core for two selected stellar  models
with $\Gamma=1.83$ and $\Gamma=2,$ respectively. 
The amplitude of the density jump is 
$\Delta\rho /\rho_{\rm d} = 0.2$.
The maximum frequency is reached when the radii of the internal core and of the outer
region are comparable. For all the studied models $\nu_{\rm{max}}$ is reached for
 $0.4 \leq R_{\rm d} /R \leq 0.6$.} \label{gr}
\end{figure}
\end{center}

\clearpage


\end{document}